\documentclass[2017, review,12pt]{elsarticle}

\usepackage{amssymb}
\usepackage{amsmath,amssymb}
\usepackage{url}
\usepackage{graphicx}
\usepackage{comment}

\begin{document}

\begin{frontmatter}



\title{
Effect of self-deflection on a totally asymmetric simple exclusion process with functions of site-assignments}


\author[RCAST]{Satori Tsuzuki}
\ead{tsuzuki@jamology.rcast.u-tokyo.ac.jp}

\author[RCAST]{Daichi Yanagisawa}
\ead{tDaichi@mail.ecc.u-tokyo.ac.jp}

\author[RCAST]{Katsuhiro Nishinari}
\ead{tknishi@mail.ecc.u-tokyo.ac.j}

\address[RCAST]{Research Center for Advanced Science and Technology, The University of Tokyo, 4-6-1, Komaba, Meguro-ku, Tokyo 153-8904, Japan}

\begin{abstract}
This study proposes a model of a totally asymmetric simple exclusion process on a single channel lane with functions of site-assignments along the pitlane. 
The system model attempts to insert a new particle to the leftmost site at a certain probability by randomly selecting one of the empty sites in the pitlane, and reserving it for the particle. 
Thereafter, the particle is directed to stop at the site only once during its travel. 
Recently, the system was determined to show a self-deflection effect, in which the site usage distribution biases spontaneously toward the leftmost site, and 
the throughput becomes maximum when the site usage distribution is slightly biased to the rightmost site, instead of being an exact uniform distribution. 
Our exact analysis describes this deflection effect and show a good agreement with simulations.
\end{abstract}


\end{frontmatter}






\section{Introduction}
The totally asymmetric simple exclusion process (TASEP) has been extensively studied in the field of non-equilibrium statistical mechanics. By employing the simple characteristic that a particle hops in a single direction with the constraint of volume exclusion effect, the TASEP has been applied to many practical problems (e.g. molecular motor transport\cite{1751-8121-48-6-065001, Denisov2015}, granular flows \cite{Neumann2009,5255048}, and traffic system \cite{PhysRevE.89.042813, 1742-5468-2017-4-043204, 1751-8121-42-44-445002, 2010arXiv1001.4124Y, doi:10.1142/S0218202515400011}). Recently, the applicability of the TASEP has been expanded by combining with the stochastic process models based on the M/M/1 queueing theories \cite{Kleinrock:1975:TVQ:1096491, Abate1987}.

This study proposes a TASEP model for the problem of flows in a single channel lane with an adjacent pitlane under the open boundary condition. 
The subject of this study is that the proposed model has function of site-assignments, according to which the system tries to insert a new particle to the leftmost site at a certain probability; if successful, the system randomly selects one of the sites in the pitlane and reserves it 
for the particle to stop at once during its travel. 
This proposed model is not only scientifically interesting but also applicable to problems in engineering because such a type of mechanism can be observed in many practical cases. 
For example, during the taxiing of an aircraft in airport ground transportations, the airport traffic controller selects one of the empty spots on the apron and assigns it to the aircraft in the rapid 
taxiway coming from the runway, and then instructs the aircraft to stop at the selected spot. 
The mechanism of aircraft taxiing is exactly the same as the applied case of the model proposed in this study. 
Other real-world examples could include the vehicular traffic in a parking lot and the problem of air boarding \cite{MIURA201744, BACHMAT2008597, doi:10.1287/opre.1080.0630, STEFFEN201264}.

Several studies have reported on the TASEP by considering the absorption lane (e.g. multiple lanes\cite{Ezaki2011PositiveCE, Ichiki2016,Verma2015} and TASEP with Langmuir Kinetics
\cite{Ichiki2016, Verma2015, PhysRevE.70.046101, doi:10.7566/JPSJ.85.044001, Yanagisawa2016, 0295-5075-107-2-20007, RePEc:wsi:ijmpcx:v:18:y:2007:i:09:p:1483-1496}). The proposed model can be said to be classified into the same category in a wider sense, expect the function of site assignment in this study is the distinguishing factor compared to the models reported in the previous studies. The major scope of the current research is to clarify the relationship between the site-assignments and system properties of the proposed model; this has not been focused on in the aforementioned studies. 

The reminder of this paper is structed as follows. Section~2 describes the details of the proposed model. Section~3 presents the investigation of dependence of system properties on each parameter through simulations. In Section~4, the deflection effect of the proposed model is described through our exact analysis. Section~5 summarizes our results and concludes this paper.

\section{Model}
A schematic view of our proposal TASEP model is depicted in Fig.\ref{fig:schemview}
. The system consists of two parallel lanes: a single channel lane and an pitlane. Each lane is composed of $L$ sites. We adopted the parallel update method in which all the particles are simultaneously updated based on the following rules:
\begin{figure*}[t]
\vspace{-2.5cm}
\begin{center}
\includegraphics[bb= 0 0 1884 1041, width=1.3\textwidth,clip]{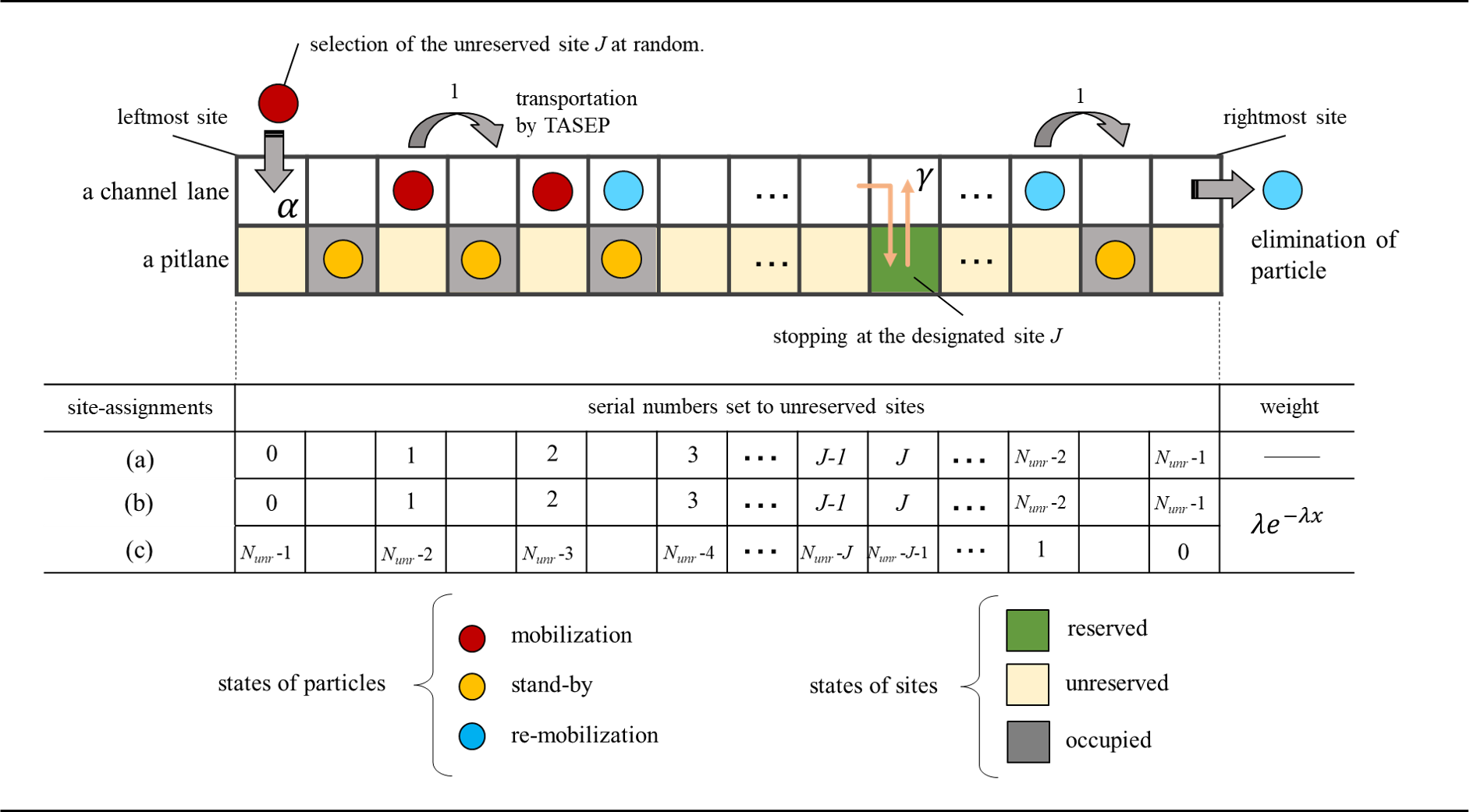}
\end{center}
\caption{Schematic view of the proposed model.}
\label{fig:schemview}
\end{figure*}

\begin{itemize}
\item{{\bf Particle Insertion}}\\
The system tries to insert a new particle to the leftmost site with the probability $\alpha$. However, if the other particle exists in the leftmost site, the attempt is blocked. In contrast, if the attempt is successful, one of the empty sites of the pitlane is reserved for the particle. Thereafter, the particle is directed to stop at the site of the pitlane only once during its travel.

\item{{\bf Site Management}}\\
The state of a site in the pitlane can be expressed in three ways: ``occupied,'' ``reserved,'' and ``unreserved.'' The ``occupied'' state indicates that the site has already stored a particle. 
The ``reserved'' state indicates that the site has already been designated for one of the particles inserted to the leftmost site. The ``unreserved'' state indicates that the site is not occupied or reserved by any particle. In site-assignments, the system sets serial numbers to unreserved sites. In the illustration shown in Fig.\ref{fig:schemview}, indices in the range of $0$ to $N_{unr}-1$ are assigned to the unreserved sites. 
The system selects one of the indices randomly, and records it to the inserted particle in the leftmost site. 

\item{{\bf Site Assignment}}\\
This study describes three methods of site-assignment: (a) uniform random distribution; (b) random distribution, which is biased by a leftmost exponential function of $\lambda e^{-\lambda x }(x \ge 0,\lambda > 0)$; and (c) random distribution, which is biased by a rightmost exponential function. 
In case of (c), the site usage distribution gets biased toward the rightmost site as parameter $\lambda$ increases 
by assigning serial numbers to the unreserved sites in the reverse order.
In this paper, we term the (b) and (c) methods simply as ``bias toward the leftmost site'' and ``bias toward the rightmost site.'' 
Note that the inverse transform sampling (ITS)\cite{doi:10.1137/1.9780898717570, Lecuyer2011} was adopted to generate biased random numbers used in (b) and (c).

\item{{\bf Transportation of Particles}}\\
The particle in the channel lane hops to the right adjacent site with a probability of $1$ in case the target site is empty. 
Here, the state of particles can be expressed in three ways: ``mobilization,'' ``stand-by,'' and ``re-mobilization.'' 
The ``mobilization'' state indicates that the particle is in transport before stopping at the designated site of the pitlane. 
The ``stand-by'' state indicates that the particle is currently stopping at the site of the pitlane. 
Futher, the ``re-mobilization'' state indicates that the particle is in transport after exiting the pitlane.
In Fig.\ref{fig:schemview}, the states are expressed as the red, yellow, and blue colored circles, respectively. 

\item{{\bf Pit-in of Particles}}\\
The particle in the channel lane hops to the lower adjacent site in the pitlane with the probability of $1$ in case the index of
the site, which is reserved to the particle, corresponds to that of the current site.

\item{{\bf Release of Particles}}\\
The particle in the pitlane hops to the upper adjacent site in the channel lane with probability $\gamma$ in case the target site in the channel lane is empty. 
The particle in the pitlane has the right of priority access against the particle in the channel lane to hop to the same site.

\item{{\bf Elimination of Particles}}\\
The particle in the channel lane is eliminated from the system at the next step after hopping to the rightmost site.
\end{itemize}

\section{Simulations}
By using the total number of time steps $N_{step}$, and total number of particles that exit from the rightmost site during 
simulation, $N_{out}$, throughput $Q$ from the rightmost site is defined as follows:
\begin{eqnarray}
Q=\frac{N_{out}}{N_{step}}
\end{eqnarray}
\subsection{The setting of input parameters}
\begin{figure}[t]
\begin{center}
\vspace{-3.0cm}
\includegraphics[width=1.33\textwidth,clip, bb= 0 0 1600 800]{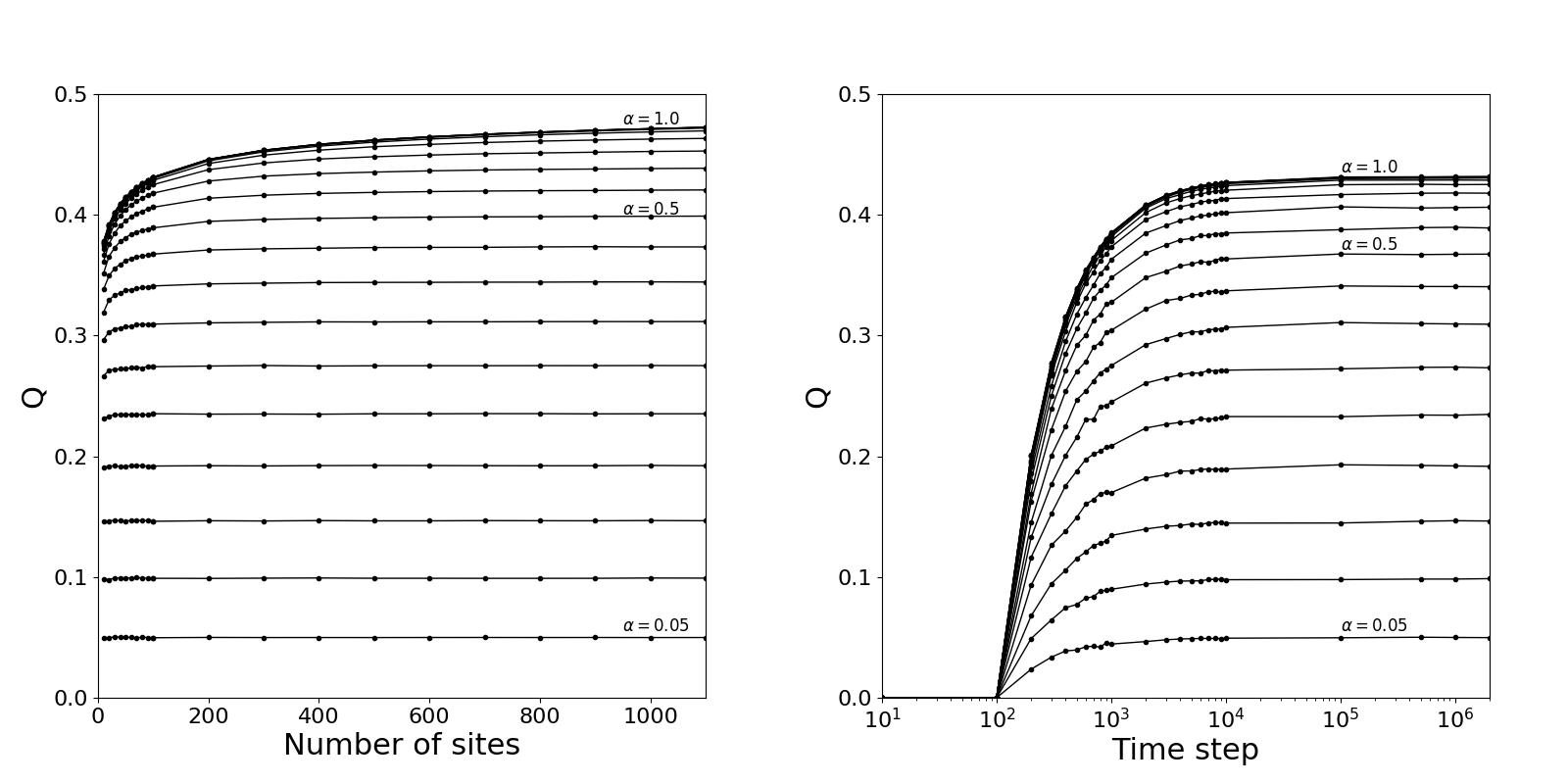}
\end{center}
\caption{Dependence of throughput $Q$ on the number of sites (left side) and number of time steps (right side) 
in case of setting the probability $\alpha$ to the different twenty values in the range of $0.05$ to $1.0$ and 
	setting the number of sites $L$ in the right side to $100$. }
\label{fig:reso-time-scaling}
\end{figure}

To determine the condition of reaching stationary state, the dependence of throughput $Q$ on each parameter is primarly 
investigated using uniform random distribution. The left and right sides in Fig.\ref{fig:reso-time-scaling} show the dependence of throughput $Q$ on 
the number of sites and number of time steps with the change in probability $\alpha$ from $0.05$ to $1.0$, respectively. 
Note that the number of sites $L$ on the right side in Fig.\ref{fig:reso-time-scaling} is fixed to $100$. 

\begin{figure}[t]
\vspace{-3.0cm}
\begin{center}
\includegraphics[width=1.33\textwidth, clip, bb= 0 0 1040 550]{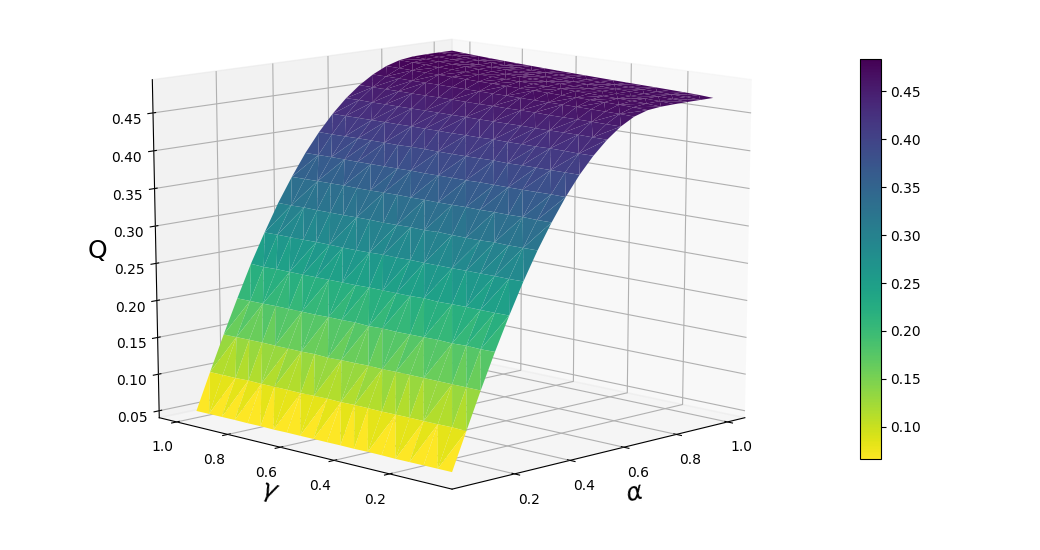}
\end{center}
\caption{Dependence of throughput $Q$ on probabilities $\alpha$ and $\gamma$.}
\label{fig:contour3D}
\end{figure}

Regarding the dependence on the number of sites, it is observed that the scalability of $Q$ slightly improves even as the number of sites increases to more than $800$. 
Regarding the dependence on time step, the scalability of $Q$ gets saturated at least from $10,0000$ times steps. 
Figure~\ref{fig:contour3D} shows the contour plot of the dependences of throughput $Q$ on probabilities $\alpha$ and $\gamma$. 
It is confirmed that throughput $Q$ turns sluggish and reaches a plateau area as probability $\alpha$ increases, whereas 
only a slight increase is observed with the increase in probability $\gamma$.

\begin{figure}[t]
\begin{center}
\vspace{-3.0cm}
\includegraphics[width=1.33\textwidth, clip, bb= 0 0 1991 950]{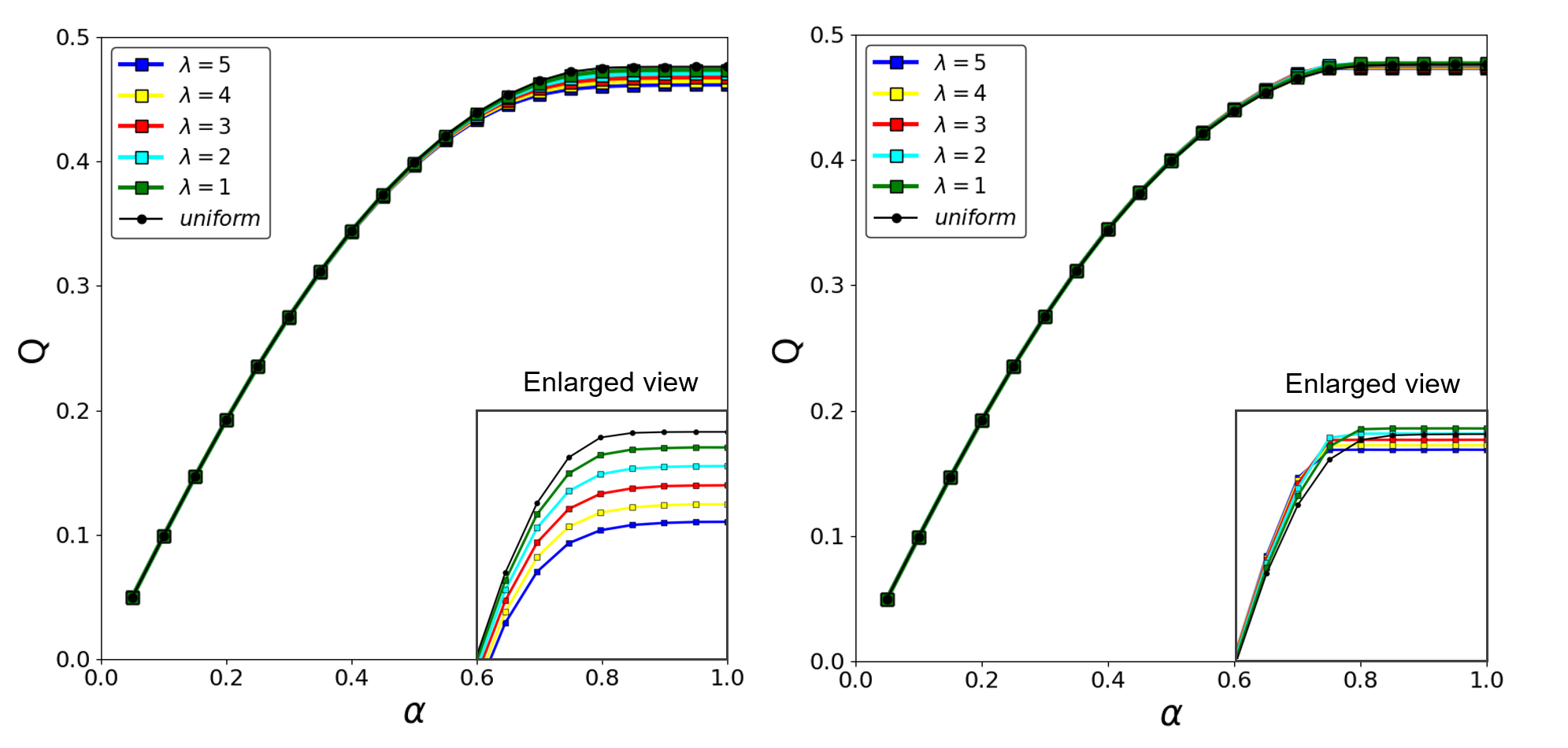}
\end{center}
\caption{Dependence of throughput $Q$ on probability $\alpha$ in case of being (b) biased toward the leftmost site (left column) and (c) biased toward the rightmost site (right column).}
\label{fig:depend-alpha}
\end{figure}

Figure~\ref{fig:depend-alpha} shows the dependence of throughput $Q$ on probability $\alpha$ in case of being (b) bias toward the leftmost site (in the left column) and (c) bias toward the rightmost site (in the right column) respectively. 
Note that the result in case of using (a) uniform random distribution is included on both sides in Fig.\ref{fig:depend-alpha} for comparison. 
The difference among simulation results was observed to be more obvious in the plateau area than the growth area~(The dependence of throughput $Q$ on the distribution parameter $\lambda$ is explained in the next section).

In response to these results, we setup the target problem as follows. The number of sites $L$ is set to be $1,000$. According to these spatial resolutions, the number of time steppings was set to be $20$ million. Probabilities $\alpha$ and $\gamma$ were set to be $1.0$ and $0.5$, respectively.

\subsection{Dependence of throughput $Q$ on parameter $\lambda$}
\begin{figure}[t]
\vspace{-3.5cm}
\begin{center}
\includegraphics[width=1.33\textwidth, clip, bb= 0 0 1280 720]{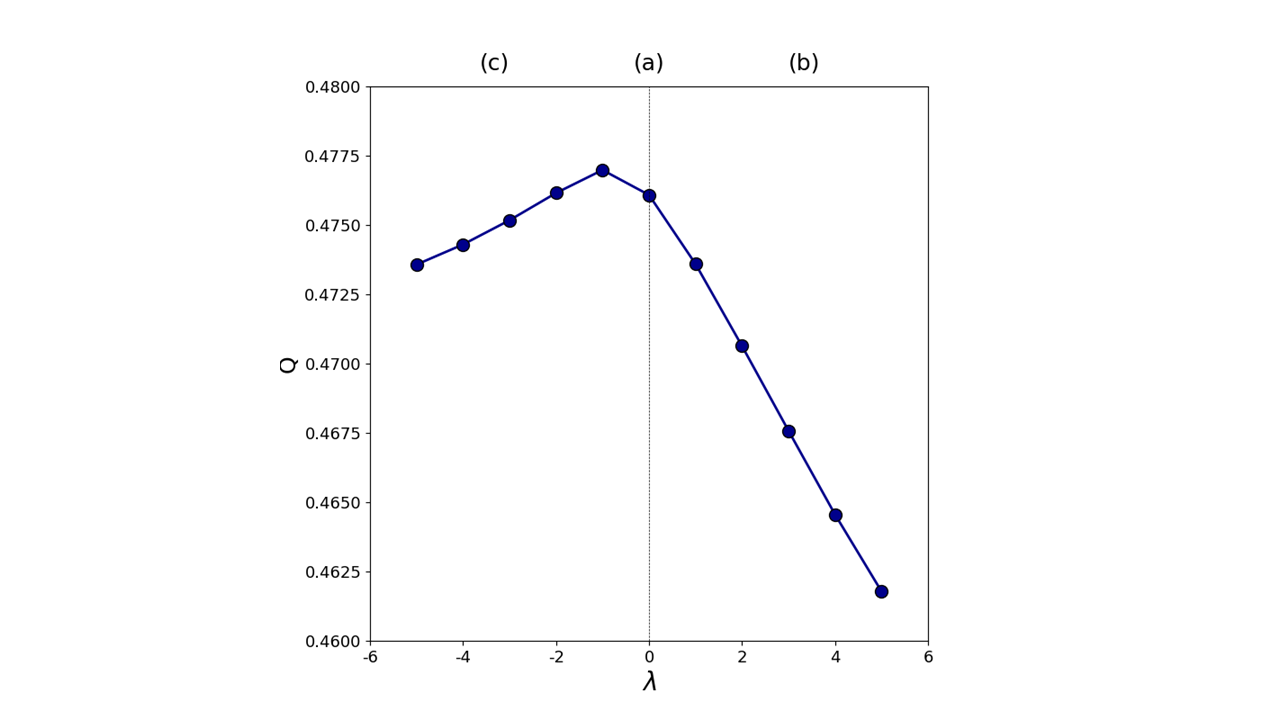}
\end{center}
\caption{Dependence of throughput $Q$ on distribution parameter $\lambda$.}
\label{fig:depend-lambda}
\end{figure}

Figure~\ref{fig:depend-lambda} shows the dependence of throughput $Q$ on distribution parameter $\lambda$ of an exponential function. 
In the case of (b) bias toward the leftmost site and (c) bias toward the rightmost site the results are displayed at the right and left of zero, respectively. 
For the sake of easy view of (c) in a single figure, parameter $\lambda$ is expressed by multiplying a negative sign in the left part, as shown in Fig.\ref{fig:depend-lambda}. 
Moreover, the result of (a) uniform random distribution is shown at zero for comparison. 
Throughput $Q$ gradually increases as parameter $\lambda$ decreases, and reaches its maximum when $\lambda = -1$, which corresponds to the case of (c), in which parameter $\lambda = 1$, and thereafter throughput $Q$ decreases. 
Here, two important observations are made. First, throughput $Q$ improves by assigning the sites closer to the rightmost site to 
particles by using (c) with $\lambda = 1$, rather than using (a) uniform random distribution; 
Second, the gradient of the line in (c) becomes moderate compared to that in (b).

To understand the reasons for these phenomena, it is reasonable to investigate the site usage distribution because the amount of usage at each site 
is equal to the occurrence count of the transport delay in this system. 
This is because the delay of particles for the traveling direction is only caused by encounters between two particles in different lanes. 
The particle in the channel lane must stop at the current site when the particle in the pitlane exits the site and hops to the adjacent right site.

\begin{figure}[t]
\begin{center}
\vspace{-3.5cm}
\includegraphics[width=1.33\textwidth, clip, bb= 0 0 2000 737]{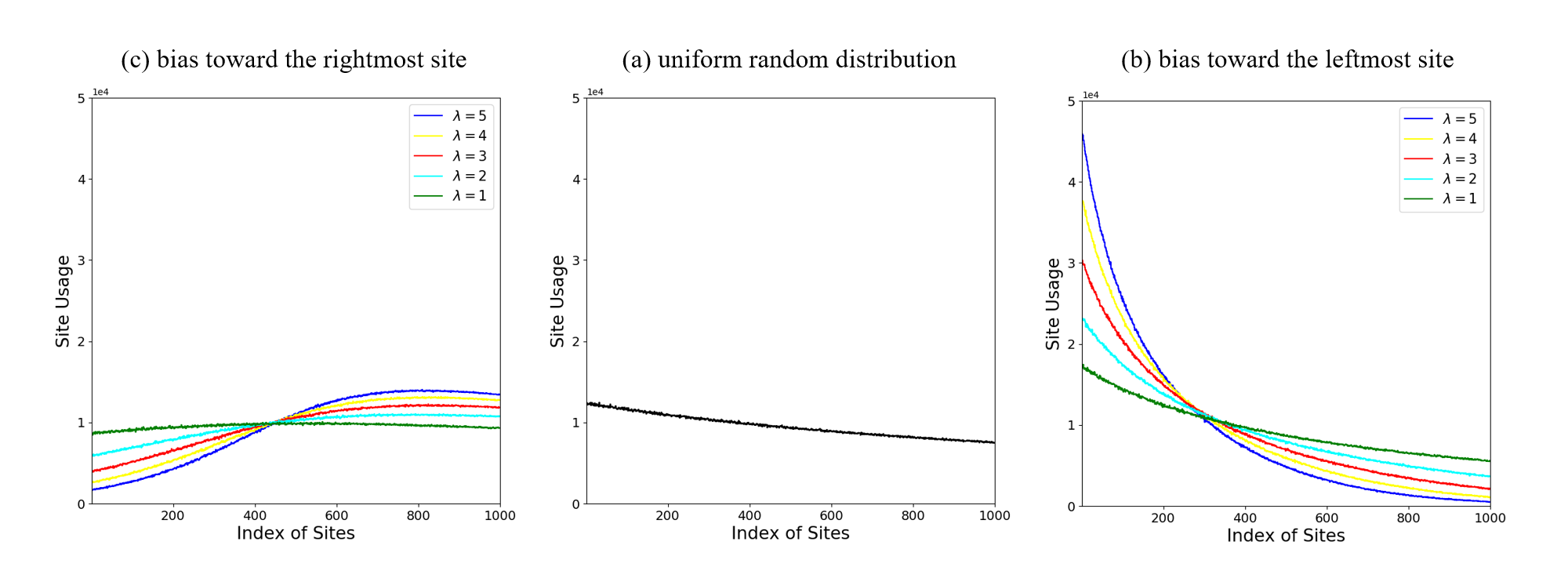}
\end{center}
\caption{Site usage distribution according to each simulation result in Fig.\ref{fig:depend-lambda}.}
\label{fig:spot-usage}
\end{figure}

Figure~\ref{fig:spot-usage} shows the site usage distribution according to each simulation result in Fig.\ref{fig:depend-lambda}. 
The results represented by each colored line correspond to those in Fig.\ref{fig:depend-alpha} represented by the same color. 
Figure~\ref{fig:spot-usage} presents an important phenomenon, which is the main topic of this paper: 
the spontaneous distribution of the site usage biases toward the leftmost site even when using (a) uniform random distribution, as shown in the center part in Fig.\ref{fig:spot-usage}.

In case of (b) bias toward the leftmost site, the site usage distribution is directly reflected by the scale of the exponential function of $\lambda e^{-\lambda x }$ according to each parameter $\lambda$. 
The traffic congestion amplified by parameter $\lambda$ provides a good explanation for 
the decline of throughput $Q$, as shown in the right part in Fig.\ref{fig:depend-lambda}.
In contrast, in case of (c) bias toward the rightmost site, the distribution shape differs from the symmetric case of (b). 
This is because each distribution is biased toward the leftmost site owing to the spontaneous bias. 
Consequently, the traffic congestion around the rightmost site is partly reduced in case of $\lambda = 1$, and therefore throughput $Q$ increases. 
Thereafter, throughput $Q$ decreases because of the traffic congestion around the rightmost site amplified by parameter $\lambda > 1$.

In the next section, we present the investigation of the cause of occurrence of the spontaneous bias toward the leftmost site and the relationship between throughput $Q$ and the spontaneous bias.

\section{Analysis}
\subsection{A four-site approximation}
This section presents the derivation of an approximation based on the Markov chain model. 
As it is difficult to consider all the $L$ sites, we approximated the proposed system by using four sites 
composed of two upper sites in the channel lane and two lower sites of the pitlane located at the right edge of two lanes.
As two kinds of particles, that is, the mobilization (red colored particles), and re-mobilization (blue colored particles) particles exist in the two sites 
in the channel lane, as mentioned in Section~2, each of the two sites in the channel lane possibly has three states: ``unoccupied,'' ``occupied by a red particle,'' and
``occupied by a blue particle.'' Moreover, each of the two sites in the pitlane has possibly two states ``occupied'' or ``unoccupied'' because only a single kind of 
particle exists in the two sites in the pitlane. Therefore, state space $S$ of this system is expressed as follows: 
\begin{eqnarray}
S=\{s(i,j,k,l)\;\;\vert\;i,j=0,1,2,k,l=0,1\}
\end{eqnarray}
Here, $s$ is a state designated by group $(i,j,k,l)$. Parameter $i$ is the serial index, which is set to the three states of 
the left site, $j$ is the serial index set to the three states of the right site in the channel lane, $k$ is the serial index set to the two states of the left site, and $l$ is the serial index set to two states of the right site 
in the pitlane. Therefore, this system has 36 possible states, provided that the rules explained in Section~2 are not considered.

Second, we define a transition probability $P^{n} (i,j,k,l)$, which is the probability that a system state at time step $n$ becomes $s(i,j,k,l)$. 
Simultaneously, we define $s_{m}^{n}$ and $P_{m}^{n}$ as abbreviations of $s(i,j,k,l)$ and $P^{n} (i,j,k,l)$, respectively. 
Here, index $m$ is a serial number assigned to all the possible states and is uniquely indicated by group $(i,j,k,l)$.

\begin{figure}[t]
\begin{center}
\vspace{-3.5cm}
\includegraphics[width=1.3\textwidth, clip, bb= 0 0 1327 1053]{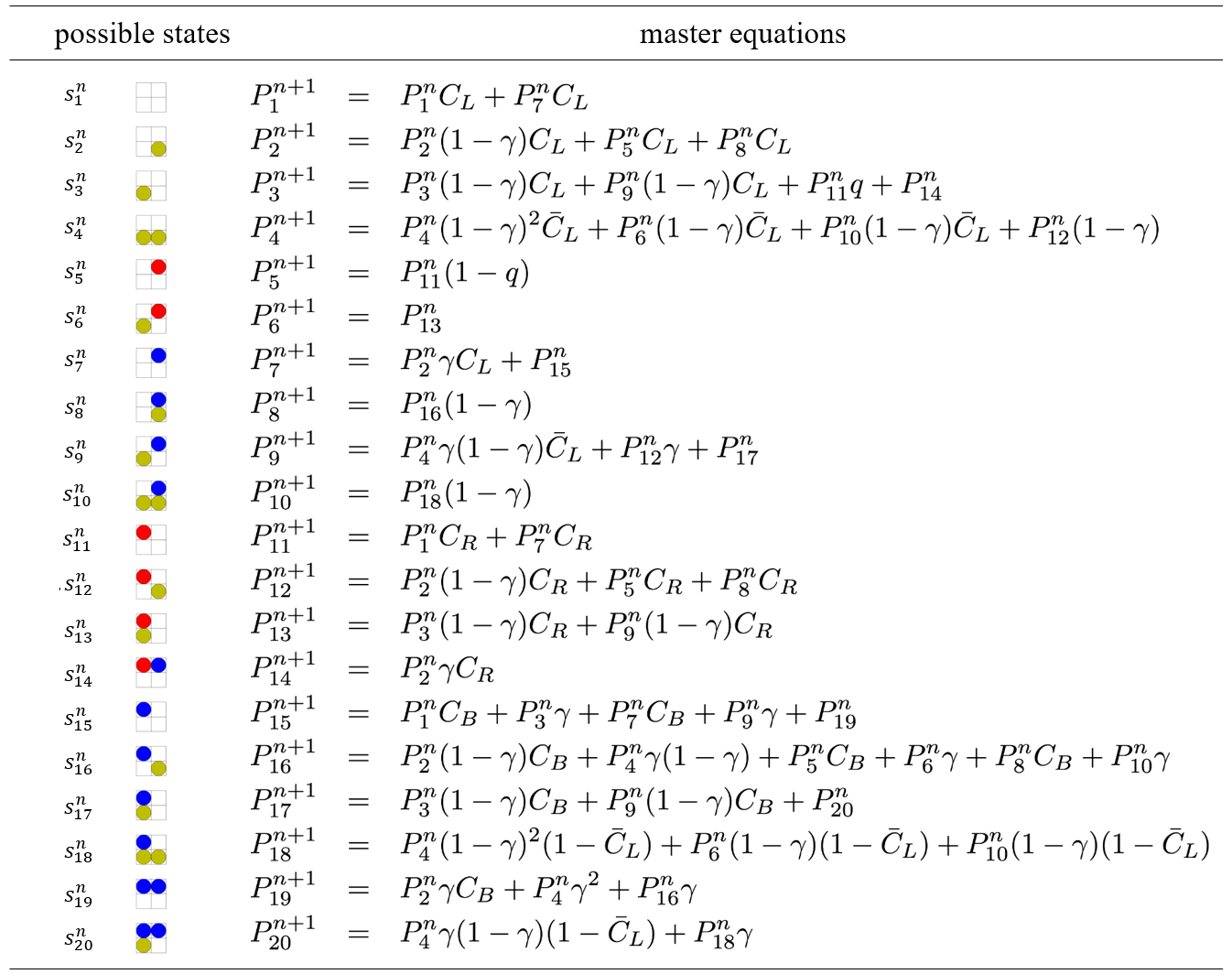}
\end{center}
\caption{The 20 possible states (left column) and master equations of the proposed system (right column).}
\label{fig:master-eq}
\end{figure}
As our point of focus was in the stationary state, we set probability $\alpha$ to $1.0$. 
Thus, by considering the rules in Section~2, the possible states are reduced from $36$ to $20$. 
Note that state space $\hat{S}$ spanned by these $20$ states is subspace of $S$. 
Figure~\ref{fig:master-eq} shows the images of all the $20$ states in the left column, and the resultant master equations obtained by 
investigating all the state transitions among these $20$ states in the right column. 
Here, $C_R$ and $C_B$ are the probabilities that the red and  blue colored particles, respectively, enter the left site of the two sites
in the channel lane, and $C_L$ is the probability that no particle enters the system. 
It is also assumed in the master equations that the probabilities of $C_R$, $C_B$, and $C_L$ become constant at the stationary state and these constant probabilities must meet the following normalization:
\begin{eqnarray}
C_R + C_B + C_L &=& 1
\end{eqnarray}
Furthermore, transition probability $P_{m}^{n} (m=1,2,3,\dots20)$ must meet the following normalization condition:
\begin{eqnarray}
\sum_{m=1}^{20} P_{m}^{n}=1
\end{eqnarray}

In transitions from $s_{4}^{n}$, $s_{6}^{n}$, and $s_{10}^{n}$, the red colored particle is not allowed to flow into the system 
as both sites in the pitlane have already been occupied; thus, we assume another pair of constant probabilities that meet the condition in which the probability of the red colored particle flowing into the system becomes zero:
\begin{eqnarray}
\overline{C_B}+\overline{C_L}=1
\end{eqnarray}

In the end, we introduce a constant probability $q$ in the transition from $s_{11}^{n}$ to $s_{3}^{n+1}$ and from
$s_{11}^{n}$ to $s_{5}^{n+1}$ in the master equations, as shown in Fig.\ref{fig:master-eq} because only these two transitions have the selectivity of sites. 
It was assumed that a red particle hops to the lower-left and lower-right sites in the pitlane with probabilities $q$ and $1-q$, respectively, in these transitions.

The conversion of transition probability matrix $\check{P}(n)$ spanned by state $P_{m}^{n}$ $(m=1,2,\dots20)$ at time step $n$ 
to matrix $\check{P}(n+1)$ at time step $n+1$ is simply written as follows:
\begin{eqnarray}
\check{P}(n+1)=M \cdot \check{P}(n) 
\end{eqnarray}
Here, $M$ is a $20\times 20$ matrix constructed using the constant coefficients of each probability in the right column in Fig.\ref{fig:master-eq}. 
At the stationary state, since the transition probability becomes constant, the system shows the following relationship:
\begin{eqnarray}
M \cdot \check{P}(n) = \check{P}(n) \label{eq:eigeneq}
\end{eqnarray}
Equation~(\ref{eq:eigeneq}) indicates that matrix $M$ has an eigenvalue of $1$ in case the system is in the stationary state. 
In this case, the normalized eigenvectors of $M$ correspond to the elements of $\check{P}(n)$. 

\subsection{Exact analysis for $L=2$}
Let us consider the case where the length of sites $L$ is set to $2$. In this case, since the system continuously tries to insert a new red particle into
the leftmost site in the channel lane with probability $1$, the constant coefficient of $C_R$ is always $1$, and both $C_B$ and $C_L$ are always $0$. 
In addition, $\overline{C_B}$ must be set to $0$ as it is not allowed for the blue colored particle to flow into the system in transitions 
from $s_{4}^{n}$, $s_{6}^{n}$, and $s_{10}^{n}$ in case the length $L$ of sites is 2. 
Under these limitations, the eigenvector of matrix $M$ is obtained as follows:
\begin{eqnarray}
\scalebox{0.7}{$\displaystyle
\begin{bmatrix}P_1^n\\P_2^n\\P_3^n\\P_4^n\\P_5^n\\P_6^n\\P_7^n\\P_8^n\\P_9^n\\P_{10}^n\\P_{11}^n\\P_{12}^n\\P_{13}^n\\P_{14}^n\\P_{15}^n\\P_{16}^n\\P_{17}^n\\P_{18}^n\\P_{19}^n\\P_{20}^n\end{bmatrix}= 35q/(334-76q) \times \begin{bmatrix}0\\0\\1\\(8-4q)/7q\\(1-q)/q\\(6q+16)/35q\\1/q\\(9-q)/35q\\(32-23q)/35q\\0\\1/q\\(44-36q)/35q\\(6q+16)/35q\\0\\1/q\\(18-2q)/(35q)\\0\\0\\(19-6q)/35q\\0\end{bmatrix}
$}
\end{eqnarray}

We estimated the expected value $E_Q$ of throughput $Q$ by summing up the state in which a blue particle exists in the upper-right site 
in the channel lane, with a weighted coefficient obtained by each transition probability. 
Considering $s_{7}^{n}$, $s_{8}^{n}$, $s_{9}^{n}$, $s_{10}^{n}$, $s_{14}^{n}$, $s_{19}^{n}$, and $s_{20}^{n}$, the expected value $E_Q$ of throughput $Q$ is obtained as follows:
\begin{eqnarray}
E_Q=\frac{95-30q}{334-76q}
\end{eqnarray}

\begin{figure}[t]
\begin{center}
\vspace{-3.5cm}
\includegraphics[width=0.9\textwidth, clip, bb= 0 0 1131 1105]{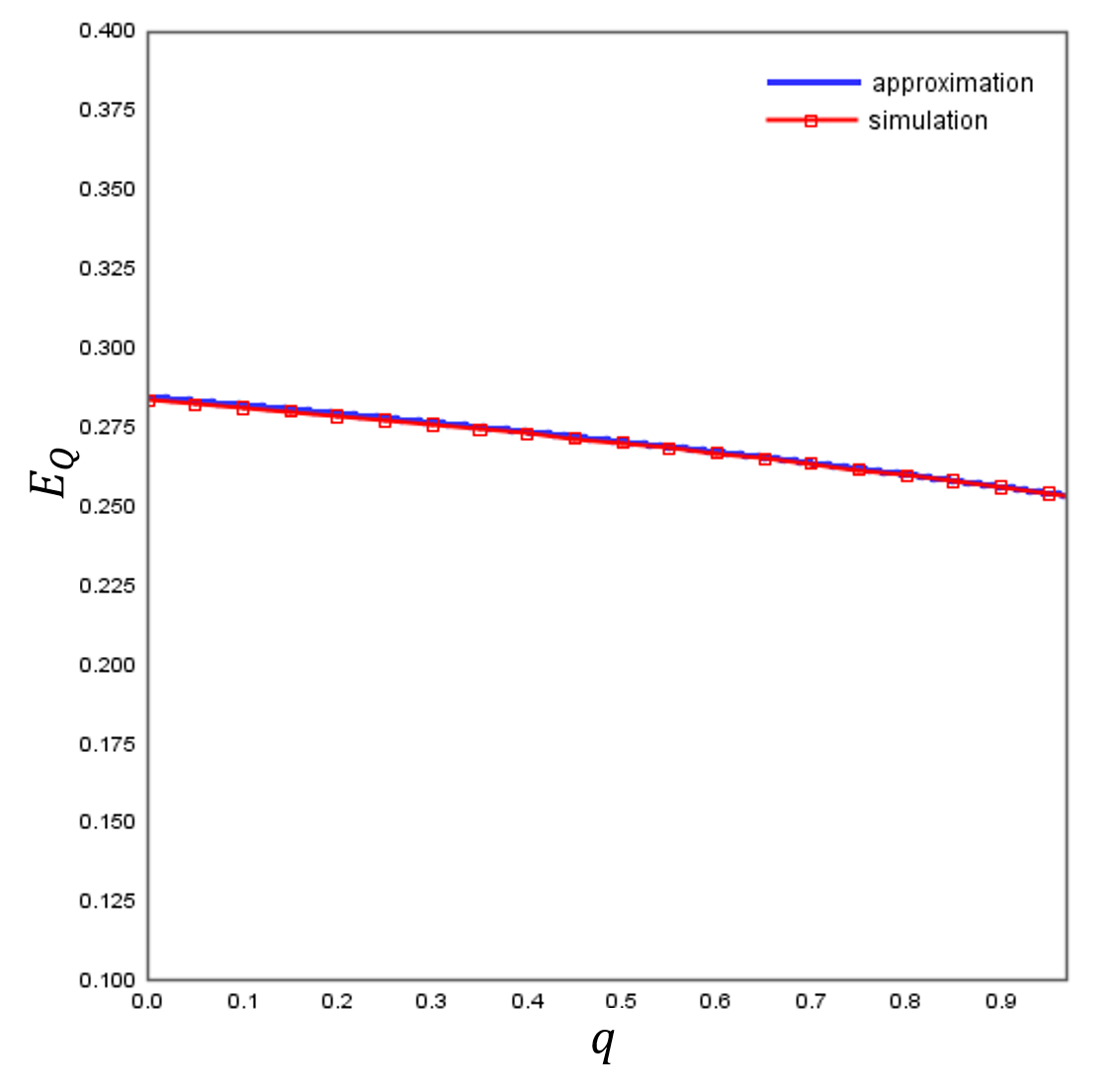}
\end{center}
\caption{Dependence of the expected throughput $E_{Q}$ on distribution probability $q$.}
\label{fig:Qappsim}
\end{figure}

The dependence of $E_Q$ of throughput $Q$ on distribution probability $q$ is shown as the blue colored line in Fig.\ref{fig:Qappsim}. 
It is found that the $E_Q$ gradually improves as $q$ decreases, implying that the throughput increases by preferentially assigning sites 
in the neighborhood of the rightmost site to the particle in the leftmost site.

The red colored line with the square symbol in Fig.\ref{fig:Qappsim} shows the dependence of throughput $Q$ on probability $q$ simulated using TASEP by 
changing probability $q$ from $0$ to $1$. 
The simulations were confirmed to show a good agreement with our exact analysis.
The fact that throughput $Q$ improves by assigning the sites in the pitlane closer to the right edge to particles is validated 
by both the exact analysis and simulations. 

Similarly, in the estimation of $E_Q$, we estimated the expected value of the usage of the left site in the pitlane by summing up the state 
in that yellow-colored particle existing in this left site, weighted by each transition probability. 
Considering $s_{3}^{n}$, $s_{4}^{n}$, $s_{6}^{n}$, $s_{9}^{n}$, $s_{10}^{n}$, $s_{13}^{n}$, $s_{17}^{n}$, $s_{18}^{n}$, and $s_{20}^{n}$, the expected value $E_S^{L}$ of the site usage of the left site in the pitlane is estimated as follows:
\begin{eqnarray}
E_S^{L}= \frac{104+4q}{334-76q}
\end{eqnarray}
Similarly, we estimated the expected value of the usage of the right site by summing up the state in which
a yellow-colored particle exists in the right site in the pitlane, weighted by each transition probability. 
Considering $s_{2}^n$,$s_{4}^n$, $s_{8}^n$, $s_{10}^n$, $s_{12}^n$, $s_{16}^n$, and $s_{18}^n$, the expected value $E_S^{R}$ of the usage of the right site in the pitlane is estimated as follows:
\begin{eqnarray}
E_S^{R}= \frac{111-59q}{334-76q}
\end{eqnarray}

\begin{figure}[t]
\begin{center}
\vspace{-3.5cm}
\includegraphics[width=0.9\textwidth, clip, bb= 0 0 1124 1124]{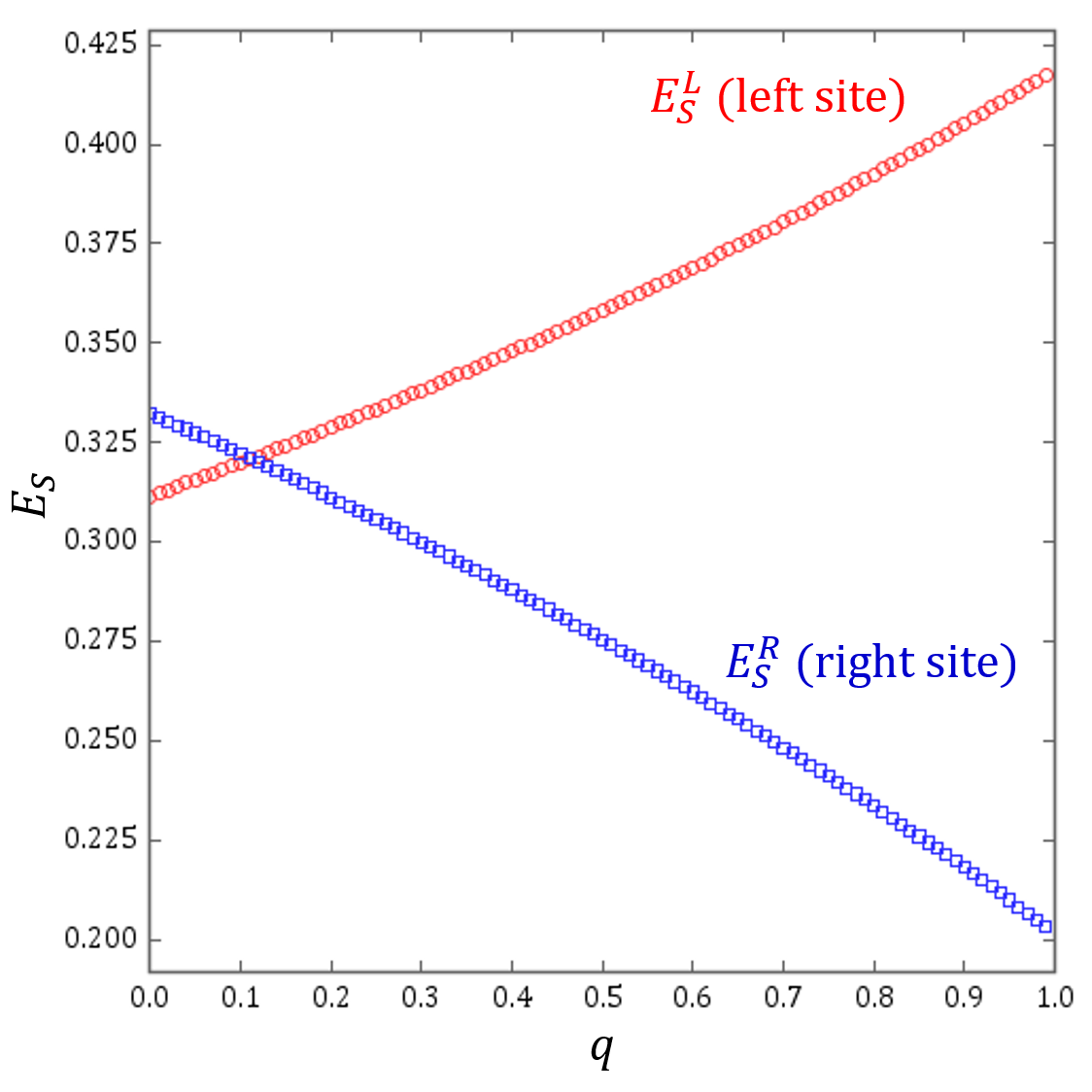}
\end{center}
\caption{Comparison of the expected value $E_S$ of the site usage for left site (red colored line) and right site (blue colored line) in the pitlane.}
\label{fig:siteusage}
\end{figure}

Figure~\ref{fig:siteusage} shows the dependence of the expected site usage $E_S^{L}$ of the left site in the pitlane (red colored line with the circle symbol), and 
the expected site usage $E_S^{R}$ of the right site in the pitlane (blue colored line with the square symbol) on probability $q$. 
The expected site usage $E_S^{L}$ of the left site was confirmed to increase more than the expected site usage $E_S^{R}$ of the right site 
if probability $q$ was set to $0.5$, indicating that the amount of site usage is spontaneously biased toward the leftmost site. We termed this as the ``self-deflection effect.'' 
In addition, the expected site usage $E_S^{L}$ of the left site was confirmed to decline as probability $q$ decreased, and become 
equal to the expected site usage $E_S^{R}$ of the right site at approximately $q=0.11$. 
Thereafter, $E_S^{R}$ of the right site increased more than $E_S^{L}$  of the left site.

The comparison with Fig.\ref{fig:Qappsim} showed that throughput still improves even after the expected site usage $E_S^{R}$ of 
the right site increases more than the expected site usage $E_S^{L}$ of the left site. 
This shows that the maximum throughput is achieved when considering the slight bias toward the rightmost site, rather than the uniform distribution.

In more detail, this deflection effect is recognized as follows: no matter how we add external forces to
change the site usage distribution biased toward the rightmost site, such external force only affects the transitions 
from $s_{11}^n$ to $s_3^{n+1}$ and from $s_{11}^n$ to $s_5^{n+1}$ in the master equations as probability $q$ only emerges in these transitions. 
Owning to the other states, the site usage distribution spontaneously biases toward the leftmost site. 
In other words, this highlights the trade-off relationship between the ``self-deflection effect'' and the bias given by external forces. 
In both exact analysis and simulations, the throughput was found to improve by the addition of external bias toward the opposite direction of the self-deflection. 

The self-deflection effect can be qualitatively explained by considering traveling times of particles as follows. 
If we set $L$ to $2$, the traveling time to the right site in the pitlane increases more than that to the left site in the pitlane;
two steps are necessary for particles to travel toward the right site in pitlane, whereas only one step is needed for them to travel toward the left site in pitlane. 
Since it is not allowed for different two particles to book the same site, the disabled time of reservations for the left site in the pitlane becomes 
shorter than that for the right site in the pitlane. 
Thus, the turnover of site usage of the left site becomes larger than that of the right site. 
Subsequently, the amount of the site usage for the left site becomes larger than that for the right site; the system spontaneously deflects. 
The relationship between the system characteristic value of throughput and site usage has not been focused on in previous studies; this is 
quantitively described by exact analysis and simulations in this paper.

The system discussed in this paper is scientifically interesting from the viewpoint of multi-particle dynamics. 
For example, it might be possible to find similar phenomena in the position-sensitive detection system of 
heavy particle beams for tumor therapies\cite{HeavyIon}. 
The particles reduce their kinetic energies in human tissues and emit pairs of annihilation gamma rays when they stop. 
By detecting these rays through positron emission tomography, it becomes possible to estimate the radioactive dose-distribution\cite{0031-9155-50-24-005, 0031-9155-53-3-002}. 
As the difference of traveling times of particles affects the detection efficiency at each position in many cases, 
the self-deflection effect might be observed in the dose-distribution. 
However, several possibilities exist observing this deflection effect in any other multi-particle system.

\section{Conclusion}
We introduced a TASEP model on a single channel lane with functions of site-assignments along the pitlane. 
In this paper, we investigated the relationship between the site-assignments and the characteristic values of the proposed system.
The contributions of this study are as follows:

The proposed system shows a ``self-deflection effect.'' The distribution of site usage biases spontaneously toward the leftmost site, and therefore throughput 
improves by the addition of the bias toward the rightmost site. The throughput becomes maximum when the site usage distribution becomes slightly biased 
toward the rightmost site, rather than the exact uniform distribution. 
These findings are validated in both our exact analysis and simulations. This effect has been newly found in this study.

As mentioned in the introduction, the major scope of this research is to clarify the relationship between the site-assignments and system properties of 
the proposed model under open boundary conditions. Accordingly, we obtained beneficial insights from the findings of this study.

\section*{Acknowledgements}
The authors thank J. Sato for insightful comments.
We would like to thank Editage (www.editage.jp) for English language editing.
This research was supported by MEXT as ``Exploratory Challenge on Post-K computer'' 
(Construction of Robust Transportation System Model and Optimization in Social System), partly 
supported by JSPS KAKENHI Grant Numbers 25287026 and 15K17583, and partly supported by ``Grant-in-Aid for JSPS Fellows'' in Japan.

\bibliographystyle{elsarticle-num} 
\bibliography{reference}

\clearpage
\end{document}